\documentstyle [12pt]{article}
\newcommand{\norm}[1]{{\protect\normalsize{#1}}}
\newcommand{\be}{\begin{equation}}
\newcommand{\ee}{\end{equation}}
\newcommand{\bea}{\begin{eqnarray}}
\newcommand{\ena}{\end{eqnarray}}
\newcommand{\sect}[1]{\setcounter{equation}{0}\section{#1}}
\newcommand{\vs}[1]{\rule[- #1 mm]{0mm}{#1 mm}}
\newcommand{\hs}[1]{\hspace{#1 mm}}

\newcommand{\half}{\mbox{$\frac{1}{2}$}}

\newcommand{\cd}{{\mbox{${\cal D}$}}}
\newcommand{\ck}{{\mbox{${\cal K}$}}}

\newcommand{\cw}{{\mbox{${\cal W}$}}}
\newcommand{\cg}{{\mbox{${\cal G}$}}}

\newcommand{\r}{l}
\newcommand{\s}{l}

\newcommand{\Z}{Z\hspace{-2mm}Z}

\newcommand{\N}{\mbox{$\underline{\mbox{N}}$}}
\newcommand{\bb}[1]{\begin{picture}(8,12)
            \put(4,3){\circle{8}}\put(3,10){\scriptsize{$ #1$}}
            \end{picture}}
\newcommand{\bm}[1]{\raisebox{-8pt}{\begin{picture}(8,16)
           \put(4,11){\circle{8}}\put(3,0){\scriptsize{$ #1$}}
            \end{picture}}}
\newcommand{\bz}{\stackrel{0}
           {\begin{picture}(8,8)\put(4,3){\circle{8}}\end{picture}}}
\newcommand{\bu}{\stackrel{1}
           {\begin{picture}(8,8)\put(4,3){\circle{8}}\end{picture}}}
\newcommand{\bd}{\stackrel{\frac{1}{2}}
           {\begin{picture}(8,8)\put(4,3){\circle{8}}\end{picture}}}
\newcommand{\bnz}{\stackrel{0}
           {\begin{picture}(8,8)\put(4,3){\circle*{8}}\end{picture}}}
\newcommand{\bnu}{\stackrel{1}
           {\begin{picture}(8,8)\put(4,3){\circle*{8}}\end{picture}}}
\newcommand{\bnd}{\stackrel{\frac{1}{2}}
           {\begin{picture}(8,8)\put(4,3){\circle*{8}}\end{picture}}}
\newcommand{\bus}{\,{\bu}\hspace{-6pt}
            \raisebox{6mm}{\begin{picture}(4,8)\thicklines
            \put(2,0){\vector(0,1){8.4}}\end{picture}}\hspace{-4pt}
            \raisebox{-2mm}{\begin{picture}(4,10)\thicklines
            \put(2,0){\vector(0,-1){12}}\end{picture}} \hspace{3pt}}

\newcommand{\tp}{\hs{-1}\rule[1mm]{1.5mm}{.1mm}\hs{-1}}
\newcommand{\tr}{\!\!-\!\!}
\newcommand{\td}{\!\!=\!}

\newcommand{\po}{\,\cdot\,}
\newcommand{\pol}{\,\cdots\,}
\newcommand{\pos}{\,\cdots\hs{-4}
            \raisebox{-1.5mm}{\begin{picture}(4,28)\thicklines
            \put(2,4){\vector(0,1){20}}\put(2,20){\vector(0,-1){20}}
            \end{picture}}\hs{4}}

\newcommand{\und}[1]{\underbrace{#1}}
\newcommand{\uo}[1]{\und{\bz\tr\po\tr\bz}_{#1}}
\newcommand{\uno}[1]{\und{\bnz\tr\po\tr\bnz}_{#1}}
\newcommand{\puo}[1]{\und{\bz\tp\po\tp\bz}_{#1}}
\newcommand{\puno}[1]{\und{\bnz\tp\po\tp\bnz}_{#1}}

\newcommand{\bq}[3]{\,\raisebox{-8pt}{\begin{picture}(15,24)(0,-8)
           \put(4,3){\circle{8}}
            \put(3,-8){\scriptsize{$ #1$}}
           \put(8,3){\line(1,0){6}}\put(18,3){\circle{8}}
            \put(24,2){\scriptsize{$ #2$}}
           \put(4,7){\line(0,1){6}}\put(4,17){\circle{8}}
            \put(11,15){\scriptsize{$ #3$}}
           \end{picture}}}

\begin{document}
\renewcommand{\thefootnote}{\fnsymbol{footnote}}
%\newpage
\pagestyle{empty}
\setcounter{page}{0}

\rightline{NBI-HE-92-93}
\rightline{December 1992}

\vs{20}

\begin{center}

{\LARGE {\bf Characteristic Dynkin Diagrams\\[4mm]
and $W$ algebras.}}\\[1.4cm]

{\large E. Ragoucy\footnote{On leave from absence of
Laboratoire de Physique Th\'eorique
{\small E}\norm{N}{\large S}{\Large L}{\large A}\norm{P}{\small P},
Chemin de Bellevue BP 110, F - 74941 Annecy-le-Vieux Cedex, France.
E-mail: RAGOUCY@FRLAPP51.IN2P3.FR}
}\\[1cm]
{\em NORDITA, Blegdamsvej 17, \\
DK-2100 Copenhagen \O, Denmark}\\[.5cm]

\end{center}

\vs{20}

\centerline{\bf Abstract}

\indent

We present a classification of characteristic Dynkin diagrams for the
$A_N$, $B_N$, $C_N$ and $D_N$ algebras. This classification is
related to the classification of \cw(\cg,\ck) algebras arising from
non-Abelian Toda models, and we argue that it can give new insight
on the structure of $W$ algebras.

\newpage
\pagestyle{plain}

\renewcommand{\thefootnote}{\arabic{footnote}}
\setcounter{footnote}{0}

\sect{Introduction \label{intro}}

\indent

These recent years, a lot of efforts have been made in order to study
the extended conformal algebras \cite{B&S}, also called $W$
algebras. Two of the
main problems concerning these algebras are that, first,
they have non linear
commutation relations, so that there are technical difficulties to deal
with, and, second, that there exists no classification
scheme, so that a global point of view for their study is hard to find.
In order to classify these algebras, one needs to determine which
features are relevant to fully distinguish their structure. Of course, the
conformal spin contents will be one of them, but it is known that it
is not sufficient to define a $W$ algebra. Some results have been
obtained in that direction \cite{nous}, for a wide class of $W$ algebras,
namely the so-called \cw(\cg,\ck) algebras that arise in the framework
of non-Abelian Toda theories \cite{ORaf}.

In this paper, we propose what could be thought of as the Dynkin diagrams
(DDs) for \cw(\cg,\ck) algebras: the Characteristic Dynkin Diagrams (CDDs).
The CDDs (introduced by Dynkin \cite{dyn}) are DDs where each node carry
a grade 0, $\half$, or 1, and are in one to one correspondence with the
$Sl(2)$ subalgebra of an algebra \cg\ (themselves being known as
classifying the \cw(\cg,\ck) algebras). Until now, only the CDDs of
exceptional algebras were classified \cite{dyn}, and we present in
the first sections of this paper a classification of CDDs for the other
algebras. Then, we will connect the CDDs to the \cw(\cg,\ck) algebras
and show how they can be of some help in the study of these algebras.
Finally, we will discuss some open problems in the conclusion.

\sect{Some basic notations and definitions}

\indent

{\em A Characteristic Dynkin Diagram (CDD)} is built from a Dynkin
Diagram (DD) of a simple Lie algebra $\cg$ in such a way that it
characterizes an $Sl(2)$ subalgebra of $\cg$. Practically,
to any node in the DD of \cg, one attaches the grade of the corresponding
simple root generator under the $Sl(2)$-Cartan generator.
It can be shown that these grades can only be 0, $\half$, or 1.
However, to any choice of grades (in \{0,1,\half\}) will not always
correspond an $Sl(2)$ algebra, since there exist some grading operators in
\cg\ that
are not Cartan generator of an $Sl(2)$ \cg-subalgebra. In other words,
among the $3^n$ possibilities of constructing a CDD from a rank $n$ algebra,
only a certain number of them will really be CDDs. In sections
\ref{sA},\ref{sBC},\ref{sD} we present a complete classification of the
CDDs for the
algebras $A_N$, $B_N$, $C_N$, and $D_N$. The CDDs of the exceptional
algebras have already being classified by Dynkin \cite{dyn},
and we will not study them here.

Let us also recall that the $Sl(2)$ subalgebras of $A_N$, $B_N$
and $C_N$ are classified by the regular subalgebras of these algebras.
In the case of $D_N$, one has to add the irregular embeddings of the
type $B_M\oplus B_{N-M}$ with $2M\neq N$. Still
for $D_N$ algebras,
some $Sl(2)$ subalgebras have a multiplicity 2 (3 for $D_4$), which
means that there are two (three) non-conjugate $Sl(2)$ related by an outer
automorphism of $D_N$ ($D_4$). Using CDDs, this multiplicity
is given by the quotient of the number of symmetries
of its CDD by the number of symmetries of the DD: thus, the multiplicity
of any $Sl(2)$ will come naturally with its CDD, and we do not have to
pay attention to it.

\indent

We will use intensively the connection that exists between CDDs and
defining vectors. A defining vector $f=(f_1,f_2,..,f_N)$
is formed by the components $f_i$ of
the $Sl(2)$-Cartan generator in a given basis. For a good choice
of basis, the grades of the simple roots (i.e. the nodes of the CDD)
can be expressed in a simple way as function of the $f_i$'s: for
any algebra \cg=A,B,C or D, and
denoting by $e_i$ a basis of $\Re^N$ where $N$=rank\cg\
(rank\cg+1 for $A_M$), the grade
of simple roots of type $e_{i+1}\pm e_i$ (when they exist in \cg)
will be $f_{i+1}\pm f_i$
and the grade of $2e_i$ and $e_i$ will be $2f_i$ and $f_i$
(respectively). With this
normalisation, the defining vector of any $Sl(2)$ can be directly
read from its CDD, and vice versa.

\indent

Finally, let us introduce some notations:

We will call {\em integral CDD}, a CDD where all the coefficients are
0 or 1. The corresponding gradation of the algebra will be integral.
Among the integral CDDs, the {\em principal} CDD is the one where all
the nodes have grade 1: it correspond to the principal $Sl(2)$ in \cg.

A {\em half-integral CDD} will be a CDD where the coefficients are only
0 or $\half$, whereas a {\em mixed CDD} will contain 0, $\half$ and 1. These
two types of CDD correspond to half-integral gradations.

We will call a {\em piece of length p} any part of a CDD formed by
$p$ nodes. Thus, a node is also a piece of length 1.

We will also draw some CDDs as
\be
\ (\,\bb{s_1}\:\tr\,)\ {\bb{s_2}}\tr\bb{s_3}\tr\bb{s_4}\tr
\pol\pol\tr\bb{s_n} \label{do}
\ee
where the notation $(\,\bb{s_1}\:\tr\ )$ means that both CDDs with and
without the node $\bb{s_1}$ have to be considered. For this kind of
CDDs, we will use a plus (+) and a minus ($-$) sign to distinguish
the CDD with the node from the other (without
the node). In the above example, these CDDs will be denoted as (\ref{do}$^+$)
and (\ref{do}$^-$).

\sect{Characteristic Dynkin Diagrams for $A_N$ \label{sA}}

\indent

Let us first remark that the CDDs of $A_N$ are symmetric. This can be proven
in two ways. First, because the $Sl(2)$-Cartan generator
can be chosen diagonal in $A_N$: as it is traceless, it can be chosen
symmetric, so that the corresponding CDD is also symmetric.
Second, because any $Sl(2)$ has a
multiplicity one: since the DD of $A_N$ possesses a $\Z_2$ symmetry, so must
any CDD. Notice that the symmetry centre of $A_N$ is either a node
(for $A_{2N+1}$), or a link between two nodes (for $A_{2N}$).
We will indicate the middle (i.e. the symmetry centre) by
a double arrow
\raisebox{-4mm}{\begin{picture}(4,24)\thicklines\put(2,4){\vector(0,1){20}}
\put(2,20){\vector(0,-1){20}}\end{picture}}, and draw only half of each CDD
of $A_N$.

\subsection{Integral CDDs of $A_N$}

\indent

Let us look first at an integral CDD. It is of the form
\be
\und{\bu\tr\po\tr\bu}_{n_1}\tr\uo{p_1}
\tr\und{\bu\tr\po\tr\bu}_{n_2}\tr\uo{p_2}\tr
\pol\pol\tr\und{\bu\tr\po\tr\bu}_{n_1}
\ee
where we can suppose $p_i\neq0$ ($\forall i$) without any loss of generality.
The gradation H associated to this CDD will be of the form
{\small
\be
(x,x-1,x-2,..,x-n_1,\und{x-n_1,..,x-n_1}_{p_1},x-n_1-1,..,x-n_1-n_2,
\und{x-n_1-n_2,..,x-n_1-n_2}_{p_2},...) \label{grad1}
\ee
}
with $x$ determined by $trH=0$. $H$ is an $Sl(2)$-Cartan generator, so that its
eigenvalues can be gathered in $\cd_j$ representations. This implies that
for any eigenvalue $y$ in (\ref{grad1}), $y-1$ must also appear with a
multiplicity which is greater than or equal to the multiplicity of $y$. For
$y=x-n_1$, this means $x-n_1-1$ must appear at least $(p_1+1)$ times, so that
$n_2=1$ and $p_2\geq p_1$. The same argument for all the eigenvalues
$y_j=x-\sum_{i=1}^jn_i$ leads us to the constraints:
\be
n_i=1,\ \mbox{ and } p_i\geq p_{i-1}\ \ \forall i\geq 2
\ee
Then, the symmetry of the CDD will lead to the two types of integral CDDs for
$A_N$:

\be
\begin{array}{l}
\und{\bu\tr\po\tr\bu}_n\tr\und{\bz\tr\po\tr\bz}_{p_1}\tr
\bu\tr\und{\bz\tr\po\tr\bz}_{p_2}\tr\bu\tr\pol\tr
\tr\bu\tr\und{\bz\tr\pos\tr\bz}_{p_\ell}
\tr\bu\tr
\pol\pol\tr\und{\bu\tr\po\tr\bu}_n \\
\ \ \ \ p_1\leq p_2\leq...\leq p_\ell
\end{array}\label{A1}
\ee

\be
\begin{array}{l}
\und{\bu\tr\po\tr\bu}_n\tr\und{\bz\tr\po\tr\bz}_{p_1}\tr
\bu\tr\und{\bz\tr\po\tr\bz}_{p_2}\tr\bu\tr\pol\tr
\bu\tr\und{\bz\tr\po\tr\bz}_{p_\ell}
\tr\bus\tr
\pol\pol\tr\und{\bu\tr\po\tr\bu}_n \\
\ \ \ p_1\leq p_2\leq...\leq p_\ell
\end{array} \label{A2}
\ee

\subsection{Half-integral CDDs of $A_N$}

\indent

For half-integral CDDs, the above reasoning is still valid, with the
difference that between the eigenvalues $y$ and $y-1$, there will occur
the eigenvalue $y-\half$, which is in a different $Sl(2)$ representation.
Instead of one series of $p_i$'s (as in the integral
case), we obtain two independent series ($p_i$ and $q_i$)
which are intertwined.
Thus, the half-integral CDDs take the form:

\be
\begin{array}{l}
\und{\bd\tr\po\tr\bd}_n\tr\und{\bz\tr\po\tr\bz}_{p_1}\tr\bd\tr
\und{\bz\tr\po\tr\bz}_{q_1}\tr\bd\tr
\und{\bz\tr\po\tr\bz}_{p_2}\tr\bd\tr
\und{\bz\tr\po\tr\bz}_{q_2}\tr\pol
\tr\bd\tr\und{\bz\tr\pos\tr\bz}_u\tr\bd\tr
\pol\pol\tr\und{\bd\tr\po\tr\bd}_n \\
p_1\leq p_2\leq...\leq p_\ell
\mbox{ and } q_1\leq q_2\leq...\leq q_{\ell'}\ ;\
u=p_\ell \mbox{ or } q_{\ell'} \mbox{ depending on }
\ell'=\ell-1 \mbox{ or }\ell
\end{array} \label{A3}
\ee
Notice that the CDD (\ref{A3}) can be split in two different CDDs,
depending on the two different values $(\ell-1)$ and
$\ell$ that $\ell'$ can take.

\subsection{Mixed CDDs for $A_N$}

\indent

The computation seems very much more complicated when looking at the
mixed CDDs. However, it can be greatly simplified if one remarks that
the pieces of CDDs that contains nodes of grade 1 must be outside of the
pieces containing nodes with grade \half. In fact, looking at a piece
of CDD which possesses a grade 1 node inside two grade \half\ nodes,
one computes:
\[
\bd\tr\uo{p}\tr\bu\tr\uo{q}\tr\bd\ \longrightarrow\
\mbox{\small
$H=(...,x,\und{x-\frac{1}{2},..,x-\frac{1}{2}}_{p+1},
\und{x-\frac{3}{2},..,x-\frac{3}{2}}_{q+1},x-2,...)$}
\]
so that the eigenvalue $x-1$ does not appear, as it should do.
Thus, the mixed CDDs are formed "outside" by an integral CDD, and
"inside" by a half-integral CDD. To study the "transition" between these
two pieces of the mixed CDD is as easy as for the preceding cases.
Altogether, the mixed CDDs of $A_N$ take the form:

\be
\begin{array}{l}
\und{\bu\tp\po\tp\bu}_n\tr\und{\bd\tp\po\tp\bd}_m
\tr\puo{p_1}\tr\bd\tr\puo{q_1}
\tr\bd\tr\puo{p_2}\tr\bd\tr\puo{q_2}
\tr\pol\tr\bd\tr\und{\bz\tp\pos\tp\bz}_u\tr\bd\tr
\pol\pol\tr\und{\bu\tp\po\tp\bu}_n \\
p_1\leq p_2\leq...\leq p_\ell
\mbox{ and } q_1\leq q_2\leq...\leq q_{\ell'}\ ;\
u=p_\ell \mbox{ or } q_{\ell'} \mbox{ depending on }
\ell'=\ell-1 \mbox{ or } \ell
\end{array} \label{A4}
\ee

\be
\begin{array}{l}
\und{\bu\tp\po\tp\bu}_n\tr\und{\bz\tp\po\tp\bz}_{p_1}\tr\bu\tr
\und{\bz\tp\po\tp\bz}_{p_2}\tp\pol\tp
\bu\tr\und{\bz\tp\po\tp\bz}_{p_{\ell+1}}\tr\bd\tr\und{\bz\tp\po\tp\bz}_{q_1}
\tr\bd\tr\und{\bz\tp\po\tp\bz}_{p_{\ell+2}}
\tr\bd\tr\und{\bz\tp\po\tp\bz}_{q_2}
\tp\pol\tp\bd\tr\und{\bz\tp\pos\tp\bz}_u\tr\bd\tp\pol \\
p_1\leq p_2\leq...\leq p_\ell\leq p_{\ell+1}\leq...\leq p_{\ell+\ell'}
\ ,\ q_1\leq q_2\leq...\leq q_{\ell''} \\
%\mbox{ and }
u=p_{\ell+\ell'} \mbox{ or } q_{\ell''} \mbox{ depending on }
\ell''=\ell'-1 \mbox{ or } \ell'
\end{array} \label{A5}
\ee

\subsection{$Sl(2)$ decomposition associated to a CDD}

\indent

In the previous section, we have used the Cartan generator of the
$Sl(2)$ subalgebra to determine all the possible CDDs.
This generator also allows us to know the $Sl(2)$ representations
that enter into the fundamental representation of $A_N$.
A $\cd_j$ representation will be characterized by the eigenvalues
$(j,j-1,...,-j)$ of the Cartan generator. We give in the following
table the $Sl(2)$ decomposition for each CDD described in the previous
section. To simplify the formulae, we define $p_0=0$ and $q_0=0$.

\be
\begin{array}{c|c}
\mbox{ CDD } & \mbox{Decomposion of the fundamental of }A_{N-1} \\
&\\ \hline &\\
(\ref{A1})
& \N=\cd_{n+\ell-1}\oplus_{i=0}^{\ell-1} (p_{i+1}-p_i)\cd_{\ell-1-i} \\
& \\
(\ref{A2})
& \N=\cd_{n+\ell-1/2} \oplus_{i=0}^{\ell-1}
(p_{i+1}-p_i)\cd_{\ell-i-1/2} \\
& \\
(\ref{A3})
& \begin{array}{l}
\N= \cd_{(\ell+\ell'+n-1)/2}\oplus
\cd_{(\ell+\ell'+n-2)/2}\oplus \\
\oplus_{i=0}^{\ell-1}
\left(p_{i+1}-p_i\right)\cd_{(\ell+\ell'-1)/2-i}
\oplus_{i=0}^{\ell'-1} (q_{i+1}-q_i)\cd_{(\ell+\ell'-2)/2-i}
\end{array} \\
& \\
(\ref{A4}) &
\N= \cd_{n+(m+\ell+\ell'-1)/2}\oplus
\cd_{(m+\ell+\ell'-2)/2}\oplus\\
(n\neq0) &
\oplus_{i=0}^{\ell-1} \left(p_{i+1}-p_i\right)
\cd_{(\ell+\ell'-1)/2-i}\oplus_{i=0}^{\ell'-1}
(q_{i+1}-q_i)\cd_{(\ell+\ell'-2)/2-i} \\
& \\
(\ref{A5}) &
\N= \cd_{n+\ell+(\ell'+\ell''-1)/2}
\oplus_{i=0}^{\ell+\ell'-1}
(p_{i+1}-p_i)\cd_{\ell+(\ell'+\ell''-1)/2-i}\oplus \\
(n+\ell\neq0) &  \oplus\cd_{(\ell'+\ell''-2)/2}
\oplus_{i=0}^{\ell''-1}(q_{i+1}-q_i)\cd_{(\ell'+\ell''-2)/2-i} \\
\end{array} \label{tA}
\ee

\indent

Let us remark that these five decompositions represent all the types
of decomposition one can obtain from the reduction of $A_N$ with respect
to any of its $Sl(2)$ subalgebras.

\sect{CDDs of $B_N$ and $C_N$ \label{sBC}}

\indent

The algebras $B_N$ and $C_N$ are irregular subalgebras of $A_M$ ones. A nice
way to see how they are embedded in $A_M$, is to use the $\Z_2$ symmetry
of $A_M$-DD. This symmetry consists in exchanging the simple root $\alpha_i$
with the simple root $\alpha_{M+1-i}$. The invariant subalgebra is built on
the generators $E_{\alpha_i}+E_{\alpha_{M+1-i}}$, so that
the DD of this subalgebra is obtained by folding the $A_M$-DD. By such a
folding, we directly get the $B_N$ and $C_N$ DDs. Here,
we will use these foldings $A_{2N}\rightarrow B_N$ and
$A_{2N-1}\rightarrow C_N$ to construct not only the DDs, but
the CDDs of $B_N$ and $C_N$ from
those of $A_M$. The folding can be extended from the DD to the CDDs of $A_M$,
because the CDDs are themselves
symmetric, so that the roots that are summed have the
same grade.
Of course, only some of the CDDs of $A_{2N}$ ($A_{2N-1}$)
will lead to CDDs of $B_N$ ($C_N$). The criteria to select them will be
the $Sl(2)$ decomposition of the fundamental of $A_M$, and we will use
intensively the table \ref{tA}. As for the $A_M$ case, the integers
$p_i$, $q_i$ and also $k_i$ will satisfy the conditions:
\be
\begin{array}{l}
p_1\leq p_2\leq p_3\leq ..\leq p_\ell\leq p_{\ell+1}\leq..
\leq p_{\ell+\ell'}\ \ \ \ p_0=0 \\
q_1\leq q_2\leq q_3\leq ...\leq q_{\ell'}\mbox{ or }q_{\ell''}
\ \ \ \ q_0=0 \\
k_1\leq k_2\leq k_3\leq ...\leq k_{\ell+\ell'+\ell''}\
\ \ \ k_0=-\half\ \mbox{ (so that }2k_0+1=0)
\end{array}\label{ord}
\ee

\subsection{CDD for $B_N$ \label{sB}}

\indent

A CDD of $A_{2N}$
will lead to a CDD of $B_N$ iff all the $\cd_j$ with $j\in\half+\Z$ have
an even multiplicity (in the fundamental of $A_{2N}$). For each decomposition
given in table \ref{tA}, we determine which are the cases where the folding
can be effectively done, and then give the folded diagram and the corresponding
decomposition of the fundamental. Of course, as we are folding $A_{2N}$, we
will consider only the CDDs with an even number of nodes, so that the center
of the CDD is always on a link. Notice that under the
folding, a CDD of the type (\ref{A3}) will behave in two different ways,
depending on whether $u$ is equal to or different from 0. If $u=0$, then the
short root of $B_N$ will have a grade \half, while for $u\neq0$ it will
possess a 0-grade.

\subsubsection{Integral CDDs}

\indent

The CDD (\ref{A1}) can always be folded in $A_{2N}$, however,
for $p_\ell=2u+2\neq0$ we get the CDD (\ref{B1})
of $B_N$, while for $p_\ell=0$ (which implies $p_i=0\ \forall i$),
we obtain the principal
CDD (\ref{B2}). The CDD (\ref{A2}) corresponds to $A_N$ with N odd, so
that we do not have to consider it here.
\be
\begin{array}{l}
\und{\bu\tr\po\tr\bu}_n\tr\und{\bz\tr\po\tr\bz}_{p_1}\tr\bu\tr
\und{\bz\tr\po\tr\bz}_{p_2}\tr\bu\tr\pol\tr
\und{\bz\tr\po\tr\bz}_{p_{\ell-1}}\tr\bu\tr\und{\bz\tr\po\tr\bz}_u
\td\bnz \\
\underline{2N+1}= \cd_{n+\ell-1}\oplus_{i=0}^{\ell-1}
(p_{i+1}-p_i)\cd_{\ell-i-1}\hs{7}p_\ell=2u+2\neq0
\end{array}\label{B1}
\ee

\be
\bu\tr\bu\tr\pol\tr\bu\tr\bu=\bnu
\ \ \ \ \underline{2N+1}=\cd_n
\label{B2}
\ee

\subsubsection{Half-integral CDDs}

\indent

The CDD (\ref{A3}) provides three types of $B_N$-CDDs, depending on
$(n+\ell+\ell'-1)$ even or odd, $n=0$ or 1 (only allowed values here)
and also\footnote{Note that $q_\ell=0$ implies $q_i=0\ \forall i$.}
$q_\ell=0$ or $q_\ell>0$.

(\ref{B3}) is obtained from (\ref{A3}) with the constraints
$n=0$, $\ell'=\ell$, $p_i=2k_i+1$, and $q_\ell=2u+2\neq0$
for (\ref{B3}$^-$); and
$n=1$, $\ell'=\ell$, $p_i=2k_i+1$, and $q_\ell=2u+2\neq0$
for (\ref{B3}$^+$).

The conditions ($n=0$, $\ell'=\ell$, $p_i=2k_i+1$, and $q_\ell=0$) and
($n=1$, $\ell'=\ell-1$, $q_i=2k_i+1$, and $p_\ell=0$) on (\ref{A3})
give the CDDs (\ref{B4}$^-$) and (\ref{B4}$^+$) respectively.

(\ref{B5}) is obtained from (\ref{A3}) with the conditions $n=0$,
$\ell'=\ell-1$, $q_i=2k_i+1$, and $p_\ell=2u+2\neq0$. We recover the CDD
(\ref{B4}$^+$)
with the conditions $n=0$, $\ell'=\ell-1$, $q_i=2k_i+1$ and $p_\ell=0$
on (\ref{A3}).

\be
\begin{array}{l}
(\,\bd\tr\,) \und{\bz\tr\po\tr\bz}_{2k_1+1}
\tr\bd\tr\und{\bz\tr\po\tr\bz}_{q_1}\tr\bd\tr
\und{\bz\tr\po\tr\bz}_{2k_2+1}\tr\bd\tr
\und{\bz\tr\po\tr\bz}_{q_2}\tr\pol\tr\bd\tr
\und{\bz\tr\po\tr\bz}_{2k_\ell+1}\tr\bd\tr
\und{\bz\tr\po\tr\bz}_u\td\bnz\  \\
\underline{2N+1}=\left\{\begin{array}{ll}
\cd_{\ell-1/2}\oplus\cd_{\ell-1}\oplus_{i=0}^{\ell-1}\left[
2\left(k_{i+1}-k_i\right)\cd_{\ell-i-1/2} \oplus
(q_{i+1}-q_i)\cd_{\ell-i-1}\right] & \mbox{without} \\
\cd_\ell\oplus\cd_{\ell-1/2}\oplus_{i=0}^{\ell-1}\left[
2\left(k_{i+1}-k_i\right) \cd_{\ell-i-1/2} \oplus
(q_{i+1}-q_i) \cd_{\ell-i-1} \right]
& \mbox{with} \bd
\end{array}\right.\\
q_\ell=2u+2
\end{array} \label{B3}
\ee

\be
\begin{array}{l}
(\,\bd\tr\,)\,\und{\bz\tr\po\tr\bz}_{2k_1+1}
\tr\bd\tr\bd\tr\und{\bz\tr\po\tr\bz}_{2k_2+1}\tr\bd\tr\bd\tr
\pol\tr\bd\tr\bd\tr
\und{\bz\tr\po\tr\bz}_{2k_\ell+1}\td\bnd\ \\
\underline{2N+1}=\left\{\begin{array}{ll}
\cd_{\ell-1/2}\oplus\cd_{\ell-1}\oplus_{i=0}^{\ell-1}
2(k_{i+1}-k_i)\cd_{\ell-i-1/2} & \mbox{without} \\
\cd_{\ell}\oplus\cd_{\ell-1/2}\oplus_{i=0}^{\ell-1}
2(k_{i+1}-k_i)\cd_{\ell-i-1/2} & \mbox{with }\bd
\end{array} \right.
\end{array} \label{B4}
\ee

\be
\begin{array}{l}
\und{\bz\tr\po\tr\bz}_{p_1}\tr\bd\tr\und{\bz\tr\po\tr\bz}_{2k_1+1}
\tr\bd\tr\und{\bz\tr\po\tr\bz}_{p_2}\tr\bd\tr
\und{\bz\tr\po\tr\bz}_{2k_2+1}\tr\pol\tr\bd\tr
\und{\bz\tr\po\tr\bz}_{2k_{\ell-1}+1}\tr\bd\tr
\und{\bz\tr\po\tr\bz}_u\td\bnz\  \\
\underline{2N+1}= \cd_{\ell-1}\oplus\cd_{\ell-3/2}
\oplus_{i=0}^{\ell-1}\left(p_{i+1}-p_i\right)\cd_{\ell-i-1}
\oplus_{i=0}^{\ell-2} (k_{i+1}-k_i)\cd_{\ell-i-3/2}
\label{B5}
\\ p_{\ell} =2u+2
\end{array}
\ee

\subsubsection{Mixed CDDs}

\indent

Studying (\ref{A4}) and (\ref{A5}) leads to the following CDDs:

(\ref{B6}) is obtained from (\ref{A4}) by the restrictions
$m=1$, $p_i=2k_i+1$,
$\ell'=\ell$ and $q_\ell=2u+2\neq0$.

(\ref{B7}) with $\ell=0$ comes from (\ref{A4}) by the conditions $m=0$,
$\ell'=\ell-1$, $q_i=2k_i+1$ and $p_\ell=2u+2\neq0$, while the general case
(for (\ref{B7})) is given by (\ref{A5}) with $\ell''=\ell'-1$, $q_i=2k_i+1$
and $p_{\ell+\ell'}=2u+2\neq0$.

(\ref{B8}) is the folding of (\ref{A5}) when applying
$\ell''=\ell'$, $n=0$, $p_i=2k_i+1$ and $q_{\ell''}=2u+2\neq0$.

(\ref{B9}) is obtained from (\ref{A5}) by the restrictions
$\ell''=\ell'$, $n=0$, $p_i=2k_i+1$ and $q_{\ell}=0$.

(\ref{B10}) can be obtained in several ways: from (\ref{A4}) with the
conditions
($\ell'=\ell-1$, $m=0$, $q_i=2k_i+1$ and $p_{\ell}=0$) or
($\ell'=\ell$, $m=1$, $p_i=2k_i+1$ and $q_{\ell}=0$), and also from
(\ref{A5}) when imposing
$\ell''=\ell'-1$, $q_i=2k_i+1$ and $p_{\ell+\ell'}=0$.

\be
\begin{array}{l}
\und{\bu\tr\po\tr\bu}_n\tr\bd\tr\uo{2k_1+1}\tr\bd\tr\uo{q_1}
\tr\bd\tr\uo{2k_2+1}\tr\bd\tr\uo{q_2}\tr\pol\tr\bd\tr\uo{2k_\ell+1}
\tr\bd\tr\uo{u}\td\bnz\\
\underline{2N+1}= \cd_{n+\ell} \oplus\cd_{\ell-1/2}
\oplus_{i=0}^{\ell-1} \left[
2\left(k_{i+1}-k_i\right)\cd_{\ell-i-1/2}
\oplus (q_{i+1}-q_i)\cd_{\ell-i-1}\right]
\\ \ \ q_{\ell}=2u+2
\end{array}\label{B6}
\ee

\be
\begin{array}{l}
\und{\bu\tp\po\tp\bu}_n\tr\puo{p_1}\tr\bu\tr\pol\tr\bu\tr\puo{p_{\ell+1}}
\tr\bd\tr\puo{2k_1+1}
\tr\bd\tr\puo{p_2}
\tr\bd\tr\pol\tr\bd\tr\puo{2k_{\ell'-1}+1}
\tr\bd\tr\puo{u}\td\bnz\\
\underline{2N+1}=\cd_{n+\ell+\ell'-1}\oplus_{i=0}^{\ell+\ell'-1}
(p_{i+1}-p_i)\cd_{\ell+\ell'-i-1}\oplus\cd_{\ell'-3/2}
\oplus_{i=0}^{\ell'-2} 2(k_{i+1}-k_i)\cd_{\ell'-i-3/2}
\\ \ p_{\ell+\ell'}=2u+2
\end{array}\label{B7}
\ee

\be
\begin{array}{l}
\und{\bz\tp\po\tp\bz}_{2k_1+1}\tr\bu\tr\und{\bz\tp\po\tp\bz}_{2k_2+1}
\tr\bu\tr\pol\tr\bu\tr\und{\bz\tp\po\tp\bz}_{2k_{\ell+1}+1}\tr\bd\tr
\und{\bz\tp\po\tp\bz}_{q_1}\tr\bd\tr\und{\bz\tp\po\tp\bz}_{2k_{\ell+2}+1}
\tr\bd\tr\und{\bz\tp\po\tp\bz}_{q_2}\tr\pol\tr
\und{\bz\tp\po\tp\bz}_{2k_{\ell+\ell'}+1}\tr\bd\tr
\und{\bz\tp\po\tp\bz}_{u}\td\bnz \\
\underline{2N+1}= \cd_{\ell+\ell'-1/2}
\oplus_{i=0}^{\ell+\ell'-1} 2(k_{i+1}-k_i)\cd_{\ell+\ell'-i-1/2}
\oplus\cd_{\ell'-1}
\oplus_{i=0}^{\ell'-1} (q_{i+1}-q_i)\cd_{\ell'-i-1}
\\ \ q_{\ell'}=2u+2
\end{array}\label{B8}
\ee

\be
\begin{array}{l}
\puo{2k_1+1}\tr\bu\tr\puo{2k_2+1}\tr\bu\tr\pol
\tr\bu\tr\puo{2k_{\ell+1}+1}
\tr\bd\tr\tr\bd\tr\puo{2k_{\ell+2}+1}\tr\pol
\tr\bd\tr\bd\tr\puo{2k_{\ell+\ell'}+1}\td\bnd \\
\underline{2N+1}= \cd_{\ell+\ell'-1/2}\oplus\cd_{\ell'-1}
\oplus_{i=0}^{\ell+\ell'-1}
2(k_{i+1}-k_i)\cd_{\ell+\ell'-i-1/2}
\end{array}\label{B9}
\ee

\be
\begin{array}{l}
\und{\bu\tr\po\tr\bu}_n\tr\bd\tr\uo{2k_1+1}
\tr\bd\tr\bd\tr\und{\bz\tr\po\tr\bz}_{2k_2+1}\tr\bd\tr\bd\tr
\pol\tr\und{\bz\tr\po\tr\bz}_{2k_{\ell-1}+1}\tr\bd\tr\bd\tr
\und{\bz\tr\po\tr\bz}_{2k_\ell+1}\td\bnd \\
\underline{2N+1}=\cd_{n+\ell}\oplus\cd_{\ell-1/2}
\oplus_{i=0}^{\ell-1} 2(k_{i+1}-k_i)\cd_{\ell-i-1/2}
\end{array}  \label{B10}
\ee

\subsection{CDD for $C_N$ \label{sC}}

\indent

We obtain the CDD for $C_N$ by folding $A_{2N-1}$. A $A_N$-CDD will
give a CDD of $C_N$ iff all the $\cd_j$ with $j\in\Z$ have
an even multiplicity (in the fundamental of $A_{2N-1}$).

\subsubsection{Integral CDDs}

\indent

We start again with the CDD (\ref{A1}). To get an even multiplicity for
$\cd_{n+\ell}$ one must impose $n=0$ and $p_1$ odd. Then, looking at the other
\cd-representations leads to the constraints $(n=0,\, p_i=2k_i+1)$, and
the folded CDD (\ref{C1}).
The CDD (\ref{A2}) can always be folded: we
obtain then the CDD (\ref{C2}).

\be
\begin{array}{l}
\uno{2k_1+1}\tr\bnu\tr\uno{2k_2+1}\tr\bnu\tr\uno{2k_3+1}\tr\pol\tr\bnu
\tr\uno{2k_{\ell-1}+1}\tr\bnu\tr\uno{k_\ell}\td\bz \\
\underline{2N}= \cd_{\ell-1}\oplus_{i=0}^{\ell-1}
2(k_{i+1}-k_i)\cd_{\ell-i-1}
\end{array}\label{C1}
\ee

\be
\begin{array}{l}
\und{\bnu\tr\po\tr\bnu}_n\tr\uno{p_1}\tr\bnu\tr\uno{p_2}
\tr\bnu\tr\pol\tr\uno{p_{\ell-1}}\tr\bnu\tr\uno{p_\ell}\td\bu \\
\underline{2N}= \cd_{n+\ell-1/2}\oplus_{i=0}^{\ell-1}
(p_{i+1}-p_i)\cd_{\ell-i-1/2}
\end{array} \label{C2}
\ee

\subsubsection{Half-integral CDDs}

\indent

Starting with (\ref{A3}), we have to know whether $n+\ell+\ell'-1$ is
even or odd. In the first case, one must impose $n=0$, so that we have
the constraints $(n=0,\, \ell'=\ell-1,\, p_i=2k_i+1)$ and obtain
(\ref{C4}$^-$). In the second case, one is left with two possibilities:
$(n=0,\, \ell'=\ell,\, q_i=2k_i+1)$ that leads to (\ref{C3}), or
$(n=1,\, \ell'=\ell-1,\, p_i=2k_i+1)$ that leads to (\ref{C4}$^+$).
\be
\begin{array}{l}
\uno{p_1}\tr\bnd\tr\uno{2k_1+1}\tr\bnd\tr\uno{p_2}
\tr\bnd\tr\uno{2k_2+1}\tr\pol
\tr\bnd\tr\uno{p_\ell}\tr\bnd\tr\uno{k_\ell}\td\bz \\
\underline{2N}= \cd_{\ell-1/2}\oplus\cd_{\ell-1}\oplus_{i=0}^{\ell-1}
\left[\left(p_{i+1}-p_i\right)\cd_{\ell-i-1/2}\oplus
2(k_{i+1}-k_i)\cd_{\ell-i-1}\right]
\end{array}
\label{C3}
\ee

\be
\begin{array}{l}
(\,\bnd\tr\,)\,\uno{2k_1+1}\tr\bnd\tr\uno{q_1}\tr\bnd\tr
\uno{2k_2+1}\tr\bnd\tr\uno{q_2}\tr\pol\tr\bnd\tr\uno{q_{\ell-1}}
\tr\bnd\tr\uno{k_\ell}\td\bz \\
\underline{2N}=\left\{\begin{array}{l}
\cd_{\ell-1/2}\oplus\cd_{\ell-1}\oplus_{i=0}^{\ell-1}
2\left(k_{i+1}-k_i\right)\cd_{\ell-i-1} \oplus_{i=0}^{\ell-2}
(q_{i+1}-q_i)\cd_{\ell-i-3/2}\ \ \mbox{with }\bd \\
\cd_{\ell-1}\oplus\cd_{\ell-3/2}\oplus_{i=0}^{\ell-1}
2\left(k_{i+1}-k_i\right)\cd_{\ell-i-1} \oplus_{i=0}^{\ell-2}
(q_{i+1}-q_i)\cd_{\ell-i-3/2}
\ \ \mbox{without}
\end{array}\right.
\end{array}\label{C4}
\ee

\subsubsection{Mixed CDDs}

\indent

For the CDD (\ref{A4}), and recalling that $n\neq0$, we have
the constraints $(m=0,\, \ell'=\ell,\, q_i=2k_i+1)$ to get (\ref{C5}),
and $(m=1,\, \ell'=\ell-1,\, p_i=2k_i+1)$ to find (\ref{C6}).

For the CDD (\ref{A5}), $(n=0,\, \ell''=\ell'-1,\, p_i=2k_i+1)$ leads
to (\ref{C7}), while $(\ell''=\ell',\, q_i=2k_i+1)$ produces (\ref{C8}).

\be
\begin{array}{l}
\und{\bnu\tr\po\tr\bnu}_n\tr\uno{p_1}\tr\bnd\tr\uno{2k_1+1}\tr\bnd\tr
\uno{p_2}\tr\bnd\tr\uno{2k_2+1}\tr\pol
\tr\bnd\tr\uno{p_\ell}\tr\bnd\tr\uno{k_\ell}\td\bz \\
\underline{2N}=\cd_{n+\ell-1/2}
\oplus\cd_{\ell-1}
\oplus_{i=0}^{\ell-1}\left[\left(p_{i+1}-p_i\right)\cd_{\ell-i-1/2}
\oplus 2(k_{i+1}-k_i)\cd_{\ell-i-1}\right] \  \
\end{array}\label{C5}
\ee

\be
\begin{array}{l}
\und{\bnu\tr\po\tr\bnu}_n\tr\bnd\tr\uno{2k_1+1}
\tr\bnd\tr\uno{q_1}\tr\bnd\tr
\uno{2k_2+1}\tr\bnd\tr\uno{q_2}\tr\pol
\tr\bnd\tr\uno{q_{\ell-1}}\tr\bnd\tr\uno{k_\ell}
\td\bz \\
\underline{2N}= \cd_{n+\ell-1/2} \oplus\cd_{\ell-1}
\oplus_{i=0}^{\ell-1} 2\left(k_{i+1}-k_i\right)\cd_{\ell-i-1}
\oplus_{i=0}^{\ell-2} (q_{i+1}-q_i)\cd_{\ell-i-3/2}
\end{array}\label{C6}
\ee

\be
\begin{array}{l}
\puno{2k_1+1}\tr\bnu\tr\puno{2k_2+1}\tr\bnu\tr\pol\tr\bnu\tr
\puno{2k_{\ell+1}+1}\tr\bnd\tr\puno{q_1}\tr\bnd\tr
\puno{2k_{\ell+2}+1}\tr\bnd\tr\puno{q_2}\tr
\pol\tr\bnd\tr\puno{q_{\ell'-1}}\tr\bnd\tr\puno{k_{\ell+\ell'}}\td\bz\\
\underline{2N}=\cd_{\ell+\ell'-1} \oplus_{i=0}^{\ell+\ell'-1}
2(k_{i+1}-k_i)\cd_{\ell+\ell'-i-1} \oplus\cd_{\ell'-3/2}
\oplus_{i=0}^{\ell'-2}2(k_{i+1}-k_i)\cd_{\ell'-i-3/2} \
\end{array}\label{C7}
\ee

\be
\begin{array}{l}
\und{\bnu\tr\po\tr\bnu}_n\tr\uno{p_1}\tr\bnu\tr\pol
\tr\bnu\tr\uno{p_{\ell+1}}\tr\bnd\tr\puno{2k_1+1}\tr\bnd\tr\pol\tr
\puno{p_{\ell+\ell'}}\tr\bnd\tr\uno{k_{\ell'}}\td\bz\ \ \ \ \\
\underline{2N}= \cd_{n+\ell+\ell'-1/2} \oplus_{i=0}^{\ell+\ell'-1}
(p_{i+1}-p_i)\cd_{\ell+\ell'-i-1/2} \oplus\cd_{\ell'-1}
\oplus_{i=0}^{\ell'-1}2(k_{i+1}-k_i)\cd_{\ell'-i-1} \
\end{array}\label{C8}
\ee

\sect{CDDs of $D_N$ \label{sD}}

\indent

The CDDs of $D_N$ are obtained thanks to the inclusion $D_N\subset B_N$.
As the dimension of the fundamental representation of $D_N$ is
the one of $B_N$ minus 1, we must select in the CDDs of $B_N$ those
which correspond to fundamental representation decomposition with at least
one $\cd_0$. Then,
from the relation between the defining vectors and the CDDs:
\[
\left.
\begin{array}{ll}
\bb{s_1}\tr\bb{s_2}\tr\pol\tr\bb{s_{n-2}}\tr\bm{s_{n-1}}\td\!\bb{s_n}
& \mbox{with } s_i=f_i-f_{i+1},\ i=1,..,n-1 \mbox{ and } s_n=f_n \\
& \\
\bb{s_1}\tr\bb{s_2}\tr\pol\tr\!\bq{s_{n-2}}{s_{n-1}}{s_n'}
& \mbox{with } s_i=f_i-f_{i+1},\ i=1,..,n-1
\mbox{ and } s_n'=f_{n-1}+f_n
\end{array}
\right\}\ \Rightarrow\ s_n'=2s_n+s_{n-1}
\]
it is easy to deduce the CDDs of $D_N$ from the CDDs of $B_N$. We recall that
the condition (\ref{ord}) is always understood.

As far as the decomposition of the fundamental is concerned, the different
cases related to the same CDD of $B_N$ will provide the same sort of
decomposition. In fact these decompositions are just the ones of $B_N$
(minus one $\cd_0$) so
that we do not re-write them in this section.

\subsection{Integral CDDs}

\indent

Let us start with the CDD (\ref{B1}). To get a $\cd_0$ representation, we
must impose $p_\ell=2u+2>p_{\ell-1}$. Then, the computation
of $s_n'$ and of the grades entering in the "tail" of the CDD
will lead to four different
cases: $u\geq2$, $u=1$, $u=0$ and $\ell>1$, or
$u=0$ and $\ell=1$. We can suppose in (\ref{B1})
that $p_i\neq0$ since $p_1=0$ can be replaced by
$n\rightarrow(n+1)$ and $p_i\rightarrow p_{i-1}\ (i>1)$.
If $u\geq2$, then $s_n'=s_{n-1}=s_{n-2}=0$ (CDD
(\ref{D1})); for
$u=1$, we get $s_n'=s_{n-1}=0$ but $s_{n-2}=1$ (CDD (\ref{D2})); finally,
$u=0$ provides
$s_n'=s_{n-1}=1$
and $s_{n-2}=1$ or 0, depending on $\ell=1$ or $\ell>1$ respectively (CDDs
(\ref{D3}) and (\ref{D4}) respectively). Note that because of the constraints
(\ref{ord}) and $2u+2>p_{\ell-1}$, $u=0$ leads to $p_i=1$, while $u=1$
imposes $p_i\leq3$.

The CDD (\ref{B2}) is of course not allowed, since it has no $\cd_0$
representation, but two other integral CDDs (\ref{D5} and \ref{D6})
are provided by mixed
CDDs: we write these CDDs here, but explain them in the section dealing
with mixed CDDs.

\be
\begin{array}{l}
\und{\bu\tr\po\tr\bu}_n\tr\uo{p_1}\tr\bu\tr\uo{p_2}\tr\pol
\tr\bu\tr\uo{p_{\ell-1}}\tr\bu\tr\uo{u-2}\tr\bq{0}{0}{0} \\
\ \ \ 5\leq p_{\ell-1}\leq 2u+1
\end{array}\label{D1}
\ee

\be
\begin{array}{l}
\und{\bu\tr\po\tr\bu}_n\tr\und{\bz\tp\bz\tp\bz}_{p_1}
\tr\bu\tr\und{\bz\tp\bz\tp\bz}_{p_2}\tr\pol
\tr\bu\tr\und{\bz\tp\bz\tp\bz}_{p_{\ell-2}}\tr\bu\tr
\und{\bz\tp\bz\tp\bz}_{p_{\ell-1}}\tr\bq{1}{0}{0} \\
\ \ 0\leq p_i\leq3
\end{array}\label{D2}
\ee

\be
\bu\tr\bu\tr\bu\tr\pol\tr\bu\tr\bq{1}{1}{1}\ \ \ \ \
%\underline{2N}=\cd_{n-1}\oplus\cd_0
\label{D3}
\ee

\be
\begin{array}{l}
\und{\bu\tr\po\tr\bu}_n\tr\bz\tr\bu\tr
\bz\tr\pol
\tr\bu\tr\bz\tr\bu\tr\bq{0}{1}{1}
\hs{14} (\ell-1)\mbox{ nodes }\bz
\end{array}\label{D4}
\ee

\be
\begin{array}{l}
\uo{2k_1+1}\tr\bu\tr\uo{2k_2+1}\tr\bu\tr\pol
\tr\uo{2k_\ell+1}\tr\bu\tr\uo{2k_{\ell+1}-1}\tr\bq{0}{0}{1}
\end{array} \label{D5}
\ee

\be
\bz\tr\bu\tr\bz\tr\bu\tr\pol\tr\bz\tr\bu\tr\bz\tr\bq{1}{0}{1}
\ \label{D6}
\ee

\subsection{Half-integral CDDs}

\indent

As for the integral case, one has to distinguish $u=0$, $u=1$ and $u\geq2$.
However, as there are two intertwined series (the odd one being never absent),
$\ell=1$ (which should be replaced here by $p_{\ell-1}=0$)
is no more particular with respect to $\ell>1$ ($p_{\ell-1}>0$).

Starting from (\ref{B3}), one obtains the CDDs (\ref{D7}), (\ref{D8}), and
(\ref{D9}) for $u\geq2$, $u=1$ and $u=0$ respectively.

(\ref{B4}) can be considered only for the case (\ref{B4}$^-$) with
$\ell=1$: it produces the
integral CDD (\ref{D5}) with $\ell=0$.

The CDD (\ref{B5}) produces (\ref{D10}), (\ref{D11}), and (\ref{D12}) in the
same way we have expressed for (\ref{B3}).

\be
\begin{array}{l}
(\,\bd\tr\,) \uo{2k_1+1}\tr\bd\tr\uo{q_1}\tr\bd\tr\uo{2k_2+1}
\tr\bd\tr\uo{q_2}\tr\pol\tr\bd\tr\uo{2k_\ell+1}\tr\bd\tr
\uo{u-2}\tr\bq{0}{0}{0} \\
\ \ \ 5\leq q_{\ell-1}\leq 2u+1
\end{array}\label{D7}
\ee

\be
\begin{array}{l}
(\,\bd\tr\,)
\uo{2k_1+1}\tr\bd\tr\und{\bz\tp\bz\tp\bz}_{q_1}\tr\bd\tr\uo{2k_2+1}
\tr\bd\tr\und{\bz\tp\bz\tp\bz}_{q_2}\tr\bd\tr\pol\tr\und{\bz\tp\bz\tp\bz}_{q_{\ell-1}}
\tr\bd\tr\uo{2k_\ell+1}\tr
\bq{\half}{0}{0} \\
0\leq q_{\ell-1}\leq3
\end{array}\ \label{D8}
\ee

\be
\begin{array}{l}
(\,\bd\tr\,) \uo{2k_1+1}\tr\bd\tr\!\und{\bz}_{q_1}\!\tr\bd\tr\uo{2k_2+1}
\tr\bd\tr\!\und{\bz}_{q_2}\!\tr\pol\tr\bd\tr\!\und{\bz}_{q_{\ell-1}}\!
\tr\bd\tr\uo{2k_\ell}\tr
\bq{0}{\half}{\half} \\
0\leq q_{\ell-1}\leq 1
\end{array}\label{D9}
\ee

\be
\begin{array}{l}
\uo{p_1}\tr\bd\tr\uo{2k_1+1}\tr\bd\tr\uo{p_2}
\tr\bd\tr\uo{2k_2+1}
\tr\bd\tr\pol\tr\bd\tr\uo{2k_{\ell-1}+1}\tr\bd\tr
\uo{u-2}\tr\bq{0}{0}{0} \\
 \ \ 5\leq p_{\ell-1}\leq 2u+1
\end{array}\label{D10}
\ee

\be
\begin{array}{l}
\und{\bz\tp\bz\tp\bz}_{p_1}\tr\bd\tr\uo{2k_1+1}
\tr\bd\tr\und{\bz\tp\bz\tp\bz}_{p_2}\tr\bd\tr
\uo{2k_2+1}\tr\pol\tr\bd\tr
\und{\bz\tp\bz\tp\bz}_{p_{\ell-1}}\tr
\bd\tr\uo{2k_{\ell-1}+1}\tr
\bq{\half}{0}{0} \\
\  0\leq p_{\ell-1}\leq3
\end{array}\ \label{D11}
\ee

\be
\begin{array}{l}
\und{\bz}_{p_1}\!\tr\bd\tr\uo{2k_1+1}\tr\bd\tr\!\und{\bz}_{p_2}\!
\tr\bd\tr\uo{2k_2+1}\tr\bd\tr\pol
\tr\!\und{\bz}_{p_{\ell-2}}\!
\tr\bd\tr\!\und{\bz}_{p_{\ell-1}}\!\tr\bd\tr\uo{2k_{\ell-1}}\tr
\bq{0}{\half}{\half} \\
0\leq p_{\ell-1}\leq 1
\end{array}\label{D12}
\ee

\subsection{Mixed CDDs}

\indent

For the three first mixed CDDs of $B_N$, the calculation is the same
as for the half-integral case. The CDD (\ref{B6})
give rise to the CDDs (\ref{D13}), (\ref{D14}), and (\ref{D15}). The CDDs
(\ref{B7}) and (\ref{B8}) produce the CDDs (\ref{D16}), (\ref{D17}), and
(\ref{D18}) for the former, and (\ref{D19}), (\ref{D20}), and (\ref{D21}) for
the latter.

The CDD (\ref{B9}) can contribute only when $\ell'=1$, which
leads to two integral CDDs for $D_N$. For these CDDs, one
has to distinguish if $k_{\ell+1}>0$ or $k_{\ell+1}=0$. In the first case,
one obtains (\ref{D5}), and in the second (\ref{D6}).

(\ref{B10}) do not contribute at all.

\be
\begin{array}{l}
\und{\bu\tr\po\tr\bu}_n\tr\bd\tr\uo{2k_1+1}\tr\bd\tr
\uo{q_1}\tr\bd\tr\uo{2k_{2}+1}
\tr\bd\tr\uo{q_2}\tr\pol\tr\bd\tr\uo{2k_{\ell}+1}
\tr\bd\tr\uo{u-2}\tr\bq{0}{0}{0}\\
\ \ \ 5\leq q_{\ell-1}\leq 2u+1
\end{array} \label{D13}
\ee

\be
\begin{array}{l}
\und{\bu\tr\po\tr\bu}_n\tr\bd\tr\uo{2k_1+1}\tr\bd\tr
\und{\bz\tp\bz\tp\bz}_{q_1}\tr\bd\tr\uo{2k_{2}+1}
\tr\bd\tr\und{\bz\tp\bz\tp\bz}_{q_2}\tr\bd\tr\pol\tr\uo{2k_{\ell-1}+1}
\tr\bd\tr\und{\bz\tp\bz\tp\bz}_{q_{\ell-1}}
\tr\bd\tr\uo{2k_\ell+1}\tr\bq{\half}{0}{0} \\
\ \  0\leq q_{\ell-1}\leq3
\end{array}\label{D14}
\ee

\be
\begin{array}{l}
\und{\bu\tr\po\tr\bu}_n\tr\bd\tr\uo{2k_1+1}\tr\bd\tr
\!\und{\bz}_{q_1}\!\tr\bd\tr\uo{2k_{2}+1}
\tr\bd\tr\!\und{\bz}_{q_2}\!\tr\bd\tr\pol\tr\uo{2k_{\ell-1}+1}
\tr\bd\tr\!\und{\bz}_{q_{\ell-1}}\!
\tr\bd\tr\uo{2k_\ell}\tr\bq{0}{\half}{\half} \\
0\leq q_{\ell-1}\leq 1
\end{array}\label{D15}
\ee

\be
\begin{array}{l}
\und{\bu\tr\po\tr\bu}_n\tr\puo{p_1}\tr\bu\tr\puo{p_2}\tr\pol
\tr\bd\tr\puo{p_{\ell+1}}\tr\bd\tr\puo{2k_1+1}\tr\bd\tr
\puo{p_{\ell+2}}
\tr\pol\tr\bd\tr\puo{2k_{\ell'}+1}
\tr\bd\tr\puo{u-2}\tr\bq{0}{0}{0} \\
\ \ 5\leq p_{\ell+\ell'-1}\leq 2u+1
\end{array}\label{D16}
\ee

\be
\begin{array}{l}
\und{\bu\tr\po\tr\bu}_n\tr\und{\bz\tp\bz\tp\bz}_{p_1}\tr\bu
\tr\und{\bz\tp\bz\tp\bz}_{p_2}\tr\pol
\tr\bd\tr\und{\bz\tp\bz\tp\bz}_{p_{\ell+1}}\tr\bd\tr\uo{2k_1+1}\tr\bd\tr
\und{\bz\tp\bz\tp\bz}_{p_{\ell+2}}\tr\bd\tr\uo{2k_{2}+1}
\tr\pol\tr\bd\tr\uo{2k_{\ell'}+1}
\tr\bq{\half}{0}{0} \\
\ \ 0\leq p_{\ell+\ell'-1}\leq3
\end{array}\label{D17}
\ee

\be
\begin{array}{l}
\und{\bu\tr\po\tr\bu}_n\tr\!\und{\bz}_{p_1}\!\tr\bu
\tr\!\und{\bz}_{p_2}\!\tr\pol
\tr\bd\tr\!\und{\bz}_{p_{\ell+1}}\!\tr\bd\tr\uo{2k_1+1}\tr\bd\tr
\!\und{\bz}_{p_{\ell+2}}\!\tr\bd\tr\uo{2k_{2}+1}
\tr\pol\tr\bd\tr\!\und{\bz}_{p_{\ell+\ell'-1}}\!\tr\bd\tr\uo{2k_{\ell'}}
\tr\bq{0}{\half}{\half} \\
0\leq p_{\ell+\ell'-1}\leq 1
\end{array}\label{D18}
\ee

\be
\begin{array}{l}
\puo{2k_1+1}\tr\bu\tr\puo{2k_2+1}\tr\pol\tr\bu\tr
\puo{2k_{\ell+1}+1}\tr\bd\tr\puo{q_1}\tr\bd\tr\puo{2k_{\ell+2}+1}
\tr\bd\tr\puo{q_2}\tr\pol\tr\bd\tr\puo{2k_{\ell+\ell'}+1}
\tr\bd\tr\puo{u-2}\tr\bq{0}{0}{0}\\
\ \ 5\leq q_{\ell'-1} \leq 2u+1
\end{array} \label{D19}
\ee

\be
\begin{array}{l}
\puo{2k_1+1}\tr\bu\tr\puo{2k_2+1}\tr\pol\tr\bu\tr
\puo{2k_{\ell+1}+1}\tr\bd\tr\und{\bz\tp\bz\tp\bz}_{q_1}\tr\bd
\tr\puo{2k_{\ell+2}+1}
\tr\bd\tr\und{\bz\tp\bz\tp\bz}_{q_2}\tr\pol
\tr\bd\tr\und{\bz\tp\bz\tp\bz}_{q_{\ell'-1}}\tr\bd\tr\puo{2k_{\ell+\ell'}+1}
\tr\bq{\half}{0}{0} \\
\ 0\leq q_{\ell'-1}\leq3
\end{array}\label{D20}
\ee

\be
\begin{array}{l}
\puo{2k_1+1}\tr\bu\tr\puo{2k_2+1}\tr\pol\tr\bu\tr
\puo{2k_{\ell+1}+1}\tr\bd\tr\!\und{\bz}_{q_1}\!\tr\bd
\tr\puo{2k_{\ell+2}+1}
\tr\bd\tr\!\und{\bz}_{q_2}\!\tr\pol
\tr\bd\tr\!\und{\bz}_{q_{\ell'-1}}\!\tr\bd\tr\puo{2k_{\ell+\ell'}}
\tr\bq{0}{\half}{\half} \\
0\leq q_{\ell'-1}\leq 1
\end{array}\label{D21}
\ee

\subsection{Folding of the CDDs of $D_N$}

\indent

In the same way that we have folded $A_M$ to get $B_N$ and $C_N$, it is
possible, using the $\Z_2$ symmetry of the "tail" of $D_N$, to fold it
and get $B_{N-1}$. Note however that the CDDs
of $D_N$ do
not possess {\em a priori} the symmetry that allows the folding, so that
we have to select the subset of symmetric CDDs, that is to discard the
CDDs (\ref{D5}) and (\ref{D6}). Moreover, the
fundamental representation $\underline{2N-1}$
of $B_{N-1}$ is obtained from the $\underline{2N}$ representation
of $D_N$ by $\underline{2N-1}=\underline{2N}-\cd_0$: we
must also select the CDDs that correspond to a decomposition with at least
one representation $\cd_0$. This prescription will suppress all the CDDs
associated with $u=0$. Then, one can
fold the remaining CDDs of $D_N$, and re-find the results given
in section \ref{sB},
the cases $u=1$ becoming a "real" case of $u\geq2$.

\indent

As an exercise, let us also compute the CDDs of $G_2$ thanks to the $\Z_3$
symmetry of $D_4$. $G_2$-CDDs are obtained from $D_4$-CDDs by identifying
its three external nodes. We have also to impose that the decomposition
of the $\underline{8}$ of $D_4$ possesses a $\cd_0$ representation, since
the fundamental of $G_2$ has dimension 7. The CDDs of $D_4$ are:

\be
\begin{array}{l}
\\
\bz\tr\bq{0}{0}{1} \hs{14} \bu\tr\bq{0}{0}{0} \hs{24}
\bu\tr\bq{1}{0}{0} \hs{14}
\bz\tr\bq{1}{1}{0} \\
\\
\bu\tr\bq{0}{1}{1} \\
\\
\bu\tr\bq{1}{1}{1} \hs{14} \bz\tr\bq{1}{0}{0} \hs{14}
\bz\tr\bq{\half}{0}{0} \hs{14}
\bd\tr\bq{0}{\half}{\half} \\
\end{array}
\ee
The first line corresponds to non $\Z_3$-symmetric CDDs. Note
however that this
line shows how the "new" outer isomorphism
(induced by the enlarged symmetry of $D_4$ w.r.t. $D_N$, $N\neq4$)
relates some of the $Sl(2)$ of $D_4$.
On the second line lies the
only CDD of $D_4$ which is $\Z_3$ symmetric but which does not possess a
$\cd_0$ representation in its fundamental representation. Thus, we fold
the CDDs of the last line and get the CDDs of $G_2$:
\be
\bnu\equiv\bu \hs{14} \bnz\equiv\bu \hs{14}
\bnz\equiv\bd \hs{14} \bnd\equiv\bz
\ee

\sect{Application to $W$ algebras}

\indent

Most (and perhaps all) the W algebras can be realized in the context of
non Abelian Toda theories. Such models are deduced from a WZW action
$S_0(g)$, based on a group G of Lie algebra \cg, by imposing some
constraints on the (chiral) Kac-Moody currents of $S_0(g)$. These
constraints are themselves induced by a gradation $H$ of \cg.
Then a Lagrange multiplier treatment and a Gauss decomposition
$g=g_+g_0g_-$ lead to a new action $S(g_0)$ that depends only on the
0-grade (w.r.t. $H$) component $g_0$. This
provides an easy-to-handle framework for the realization of W algebras
(which are symmetry of non Abelian Toda theories) in term of the
Kac-Moody generators on the one hand, and in terms of free fields
on the other hand. The W algebra obtained is naturally associated to
an $Sl(2)$ \cg-subalgebra, $H$ being the Cartan generator of this $Sl(2)$
subalgebra. In fact, the $Sl(2)$ embeddings in \cg\
classify the different $W$ algebras which can be obtained by such
a procedure. These algebras are denoted \cw(\cg,\ck), where \ck\ is
the \cg-subalgebra whose principal $Sl(2)$ is the $Sl(2)$ algebra
under consideration. Since these $W$ algebras are classified by the
different $Sl(2)$ in \cg, it is natural to relate also these algebras
to CDDs. Let us immediately remark that such an approach will be
particularly useful for the folding of $W$ algebras \cite{fold}, since
the CDDs of $B_N$ and $C_N$ already have been construct using this
method.
We give below other examples of the use of CDDs in the study
of \cw(\cg,\ck) algebras.

\subsection{Characterization of \cw(\cg,\ck)}

\indent

As explain above, since the \cw(\cg,\ck) are classified by $Sl(2)$
subalgebras in \cg, we can think of the \cg-CDDs as the Dynkin
Diagrams for the \cw(\cg,\ck) algebras. We want here to show how the
use of the CDDs, combined with the results obtained in \cite{nous} give
a lot of informations on the structure of \cw(\cg,\ck) with a minimal
of calculation.
Using only the decomposition of the fundamental
\cg-representation with respect to this $Sl(2)$, \cite{nous} were able to
compute the spin contents of all the \cw(\cg,\ck) algebras. Moreover,
extending the $Sl(2)$ algebra to an $Sl(2)\oplus U(1)$ algebra, they
also give the eigenvalues of all the $W$ generators with respect to
some special U(1). From this work, and the knowledge of the fundamental
representation decomposition, it is possible to make contact between
CDDs and \cw(\cg,\ck) algebras. Moreover, the knowledge of the grades of
the simple roots allows us to compute the Kac-Moody subalgebra which
is in \cw(\cg,\ck), up to $U(1)$ factors. In fact, this Kac-Moody algebra
is just the "affinization" of the algebra one obtains by removing in the
CDD all the nodes that have not a 0-grade. Note that the $U(1)$ factors
that the CDD cannot determine are just the ones whose eigenvalues are
computed in \cite{nous}, so that combining the two approaches gives almost
all the informations on \cw(\cg,\ck). The results are given in the tables
below. We will use the notations
\be
\begin{array}{l}
r_i=p_{i+1}-p_i\ \ i\geq1\ \ \mbox{ and }\ r_0=p_1 \\
s_i=q_{i+1}-q_i\ \ i\geq1\ \ \mbox{ and }\ s_0=q_1 \\
\r_i=k_{i+1}-k_i\ \ i\geq1 %\ \ \ \mbox{ if } p_i=2k_i+1\\
%\s_i=k_{i+1}-k_i\ \ i\geq1\ \ \ \mbox{ if } q_i=2k_i+1
\end{array}
\ee

\subsubsection{\cw($A_N$,\ck) algebras}

\indent

The only regular subalgebras of $A_N$ are (sum of) $A_M$ subalgebras. When
reducing the fundamental representation with respect to the principal
$Sl(2)$ of $A_M$, a $\cd_{M/2}$ representation will appear. Thus,
we identify the form of \ck\ thanks to the rule
\be
\cd_{p/2}\mbox{ in fundamental}\ \longrightarrow\ A_{p}
\mbox{ in \ck}
\ee
Looking at all the CDDs of $A_N$ we obtain the following table

{\small
\be
\begin{array}{|c|c|c|}
\hline && \\
\mbox{CDD} & \ck \mbox{ in }\cw(A_N,\ck)
& \mbox{KM subalg. in \cw\ } (\mbox{up to }U(1)\mbox{ factors}) \\
&& \\ \hline && \\
(\ref{A1}) & A_{2n+2\ell-2}\oplus_{i=0}^{\ell-2}
r_iA_{2\ell-2i-2}
& A_{p_{\ell}}\oplus_{i=1}^{\ell-1}2A_{p_i} \\
&& \\
(\ref{A2}) & A_{2n+2\ell-1}\oplus_{i=0}^{\ell-2}
r_iA_{2\ell-2i-1}
& 2\oplus_{i=1}^{\ell} A_{p_i}\\
&& \\
(\ref{A3}) & A_{n+\ell+\ell'-1}\oplus A_{n+\ell+\ell'-2} \oplus
& 2\oplus_{i=1}^{\ell-1} \left[ A_{p_i}\oplus A_{q_i} \right]
\oplus A_{p_{\ell}} \ (\mbox{if }\ell'=\ell-1)\\
& \oplus_{i=0}^{\ell-1}
r_i A_{\ell+\ell'-2i-1}
\oplus_{i=0}^{\ell'-1} s_i A_{\ell+\ell'-2i-2}
& 2\oplus_{i=1}^{\ell-1} \left[ A_{p_i}\oplus A_{q_i} \right]
\oplus 2A_{p_{\ell}} \oplus A_{q_\ell}\ (\mbox{if }\ell'=\ell)\\
&& \\
(\ref{A4}) & A_{2n+m+\ell+\ell'-1}\oplus A_{m+\ell+\ell'-2} \oplus
& 2\oplus_{i=1}^{\ell-1} \left[ A_{p_i}\oplus A_{q_i} \right]
\oplus A_{p_{\ell}} \ (\mbox{if }\ell'=\ell-1)\\
(n\neq0) & \oplus_{i=0}^{\ell-1}
r_i A_{\ell+\ell'-2i-1}
\oplus_{i=0}^{\ell'-1} s_i A_{\ell+\ell'-2i-2}
& 2\oplus_{i=1}^{\ell-1} \left[ A_{p_i}\oplus A_{q_i} \right]
\oplus 2A_{p_{\ell}} \oplus A_{q_\ell}\ (\mbox{if }\ell'=\ell)\\
&& \\
(\ref{A5}) & A_{2n+2\ell+\ell'+\ell''-1}
\oplus_{i=0}^{\ell+\ell'-1}
r_i A_{2\ell+\ell'+\ell''-2i-1}
& 2\left[\oplus_{i=1}^{\ell+\ell'-1} A_{p_i}
\oplus_{i=1}^{\ell''-1} A_{q_i} \right]
\oplus A_{p_{\ell+\ell'}} \ (\mbox{if }\ell''=\ell'-1)\\
(n+\ell\neq0) & \oplus A_{\ell+\ell'-2}
\oplus_{i=0}^{\ell''-1} s_i A_{\ell+\ell'-2i-2}
& 2\oplus_{i=1}^{\ell+\ell'-1} \left[ A_{p_i}\oplus A_{q_i} \right]
\oplus 2A_{p_{\ell+\ell'}} \oplus A_{q_{\ell''}}\ (\mbox{if }\ell''=\ell')\\
&& \\ \hline
\end{array}
\ee
}

The spin contents and the hypercharges are then easily computed from the
decompositions given in table \ref{tA}, following the scheme developed
in \cite{nous}.

\subsubsection{\cw($B_N$,\ck) algebras}

\indent

$B_N$ contains as regular subalgebras $A_M$, $B_M$ and $D_M$.
The relation between these subalgebras and the decomposition of the
fundamental with respect to their principal $Sl(2)$ have been given
in \cite{nous}: each $A_M$ contributes to the fundamental
decomposition by a term $2\cd_{M/2}$, and $B_M$ or $D_M$ provide
a $\cd_M$ representation (we use the notation $B_1$ for the $A_1$
algebra of index 2, constructed on a short root). Moreover,
in $B_N$, the principal
$Sl(2)$ constructed on a $D_M$ algebra is the same as the one
constructed on a $B_M$ algebra, so that we can neglect these
inclusions. Thus, the rules are in this case
\be
\left\{
\begin{array}{l}
\cd_p\ \longrightarrow\ B_p \\
2\cd_{p/2}\ \longrightarrow\ A_{p-1}
\end{array} \right.
\ee

{\small
\be
\begin{array}{|c|c|c|}
\hline && \\
\mbox{CDD} & \ck \mbox{ in }\cw(B_N,\ck)
& \mbox{KM subalg. in \cw\ } (\mbox{up to }U(1)\mbox{ factors}) \\
&& \\ \hline && \\
(\ref{B1}) & B_{n+\ell-1}\oplus p_1B_{\ell-1}
\oplus_{i=1}^{\ell-2} r_i B_{\ell-i-1}
& B_{u+1}\oplus_{i=1}^{\ell-1}A_{p_i} \\
&& \\
(\ref{B2}) & B_N & \emptyset  \\
&& \\
(\ref{B3}^-)
& \begin{array}{l}
(k_1+1)A_{2\ell-1} \oplus (q_1+1)B_{\ell-1}\oplus \\
\oplus_{i=1}^{\ell-1} \r_i A_{2\ell-2i-1}
\oplus_{i=1}^{\ell-2} s_i B_{\ell-i-1}
\end{array}
& B_{u+1}\oplus A_{2k_\ell+1}
\oplus_{i=1}^{\ell-1} [A_{2k_i+1} \oplus A_{q_i}] \\
&& \\
(\ref{B3}^+) & \begin{array}{l}
(k_1+1)A_{2\ell-1} \oplus (q_1+1)B_{\ell}\oplus \\
\oplus_{i=1}^{\ell-1} \r_i A_{2\ell-2i-1}
\oplus_{i=1}^{\ell-2} s_i B_{\ell-i-1}
\end{array}
& B_{u+1}\oplus A_{2k_\ell+1}
\oplus_{i=1}^{\ell-1} [A_{2k_i+1} \oplus A_{q_i}] \\
&& \\
(\ref{B4}^-)
& \begin{array}{l}
(k_1+1)A_{2\ell-1} \oplus (q_1+1)B_{\ell-1}\oplus \\
\oplus_{i=1}^{\ell-1} \r_i A_{2\ell-2i-1}
\end{array}
& \oplus_{i=1}^{\ell} A_{2k_i+1} \\
&& \\
(\ref{B4}^+) & \begin{array}{l}
(k_1+1)A_{2\ell-1} \oplus (q_1+1)B_{\ell}\oplus \\
\oplus_{i=1}^{\ell-1} \r_i A_{2\ell-2i-1}
\end{array}
& \oplus_{i=1}^{\ell} A_{2k_i+1} \\
&& \\
(\ref{B5}) & \begin{array}{l}
(p_1+1)B_{\ell-1} \oplus (k_1+1)A_{2\ell-3} \oplus \\
\oplus_{i=1}^{\ell-2} [r_i B_{\ell-i-1} \oplus \s_iA_{2\ell-2i-3}]
\end{array}
& B_{u+1} \oplus_{i=1}^{\ell-1}[ A_{p_i} \oplus A_{2k_i+1}] \\
&& \\
(\ref{B6}) & \begin{array}{l}
B_{n+\ell}\oplus (k_1+1)A_{2\ell-1} \oplus q_1B_{\ell}\oplus\\
\oplus_{i=1}^{\ell-1} \r_i A_{2\ell-2i-1}
\oplus_{i=1}^{\ell-2} s_iB_{\ell-i-1}
\end{array}
& A_{2k_{\ell}+1} \oplus B_{u+1}
\oplus_{i=1}^{\ell-1} [A_{2k_i+1} \oplus A_{q_i}] \\
&& \\
(\ref{B7}) & \begin{array}{l}
B_{n+\ell+\ell'-1} \oplus q_1B_{\ell+\ell'-1}
\oplus_{i=1}^{\ell+\ell'-2} r_i B_{\ell+\ell'-i-1} \\
\oplus (k_1+1)A_{2\ell'-3} \oplus_{i=0}^{\ell'-2} \s_iA_{2\ell'-2i-3}
\end{array}
& B_{u+1} \oplus_{i=1}^{\ell+\ell'-1} A_{p_i}
\oplus_{i=1}^{\ell'-1} A_{2k_i+1} \\
&& \\
(\ref{B8}) & \begin{array}{l}
(k_1+1)A_{2\ell+2\ell'-1}
\oplus_{i=1}^{\ell+\ell'-1} \r_i A_{2\ell+2\ell'-2i-1} \oplus \\
\oplus (q_1+1)B_{\ell'-1} \oplus_{i=1}^{\ell'-2} s_i B_{\ell'-i-1} \end{array}
& B_{u+1} \oplus_{i=1}^{\ell+\ell'} A_{2k_i+1}
\oplus_{i=1}^{\ell'-1} A_{q_i} \\
&& \\
(\ref{B9}) & \begin{array}{l}
(k_1+1)A_{2\ell+2\ell'-1}\oplus B_{\ell'-1} \oplus \\
\oplus_{i=1}^{\ell+\ell'-1} \r_i A_{2\ell+2\ell'-2i-1}
\end{array}
& \oplus_{i=1}^{\ell+\ell'} A_{2k_i+1}\\
&& \\
(\ref{B10}) & B_{n+\ell}\oplus (k_1+1)A_{2\ell-1}
\oplus_{i=1}^{\ell+\ell'-1} \r_i A_{2\ell-2i-1}
& \oplus_{i=1}^{\ell} A_{2k_i+1}\\
&& \\ \hline
\end{array}
\ee
}

\subsubsection{\cw($C_N$,\ck) algebras}

\indent

As in the previous case, we can write $C_1$ to distinguish the
index 1 $A_1$ algebra constructed on a long root, from the index 2
$A_M$ subalgebras. Then, the
correspondence between $\cd_M$ representations and (regular)
subalgebras will be:
\be
\left\{
\begin{array}{l}
\cd_{p-1/2}\ \longrightarrow\ C_p \\
2\cd_{p/2}\ \longrightarrow\ A_p
\end{array} \right.
\ee

{\small
\be
\begin{array}{|c|c|c|}
\hline && \\
\mbox{CDD} & \ck \mbox{ in }\cw(C_N,\ck)
& \mbox{KM subalg. in \cw\ } (\mbox{up to }U(1)\mbox{ factors}) \\
&& \\ \hline && \\
(\ref{C1}) & (k_1+1)A_{2\ell-2} \oplus_{i=1}^{\ell-2} \r_i A_{2\ell-2i-2}
& C_{k_{\ell}+1}\oplus_{i=1}^{\ell-1} A_{2k_i+1} \\
&& \\
(\ref{C2}) & (p_1+1)C_{n+\ell} \oplus_{i=1}^{\ell-2} r_i C_{\ell-i}
& \oplus_{i=1}^{\ell} A_{p_i}\\
&& \\
(\ref{C3}) & \begin{array}{l}
(p_1+1)C_\ell\oplus (k_1+1)A_{2\ell-2} \oplus \\
\oplus_{i=1}^{\ell-2} [r_iC_{\ell-i} \oplus \s_i A_{2\ell-2i-2}]
\end{array}
& \begin{array}{l}
\oplus_{i=1}^{\ell-1} \left[ A_{p_i}\oplus A_{2k_i+1} \right] \\
\oplus A_{p_{\ell}} \oplus C_{k_\ell+1}
\end{array} \\
&& \\
(\ref{C4}^-)
& \begin{array}{l}
(k_1+1)A_{2\ell-2} \oplus (q_1+1)C_{\ell-1} \oplus \\
\oplus_{i=1}^{\ell-2} [\r_i A_{2\ell-2i-2} \oplus s_iC_{\ell-i-1}]
\end{array}
& \oplus_{i=1}^{\ell-1} \left[ A_{2k_i+1}\oplus A_{q_i}\right]
\oplus C_{k_\ell+1} \\
&& \\
(\ref{C4}^+) & \begin{array}{l}
(k_1+1)A_{2\ell-2} \oplus (q_1+1)C_{\ell} \oplus \\
\oplus_{i=1}^{\ell-2} [\r_i A_{2\ell-2i-2} \oplus s_iC_{\ell-i-1}]
\end{array}
& \oplus_{i=1}^{\ell-1} \left[ A_{2k_i+1}\oplus A_{q_i}\right]
\oplus C_{k_\ell+1} \\
&& \\
(\ref{C5}) & \begin{array}{l}
C_{n+\ell} \oplus (k_1+1)A_{2\ell-2} \oplus q_1 C_{\ell}\oplus\\
\oplus_{i=1}^{\ell-2} [r_iC_{\ell-i-1} \oplus \s_i A_{2\ell-2i-2}]
\end{array}
& \oplus_{i=1}^{\ell} A_{p_i} \oplus_{i=1}^{\ell-1} A_{2k_i+1}
\oplus C_{k_\ell+1} \\
&& \\
(\ref{C6}) & \begin{array}{l}
C_{n+\ell}\oplus (k_1+1)A_{2\ell-2}\oplus \\
\oplus_{i=1}^{\ell-2} [\r_i A_{2\ell-2i-2} \oplus s_i C_{\ell-i-1}]
\end{array}
& \oplus_{i=1}^{\ell-1} [A_{2k_i+1} \oplus A_{q_i}]
\oplus C_{k_{\ell}+1} \\
&& \\
(\ref{C7}) & \begin{array}{l}
(k_1+1)A_{2\ell+2\ell'-2}
\oplus_{i=1}^{\ell+\ell'-2} \r_i A_{2\ell+2\ell'-2i-2} \oplus \\
\oplus (q_1+1)C_{\ell'-1} \oplus_{i=1}^{\ell'-2} s_iC_{\ell'-i-1}
\end{array}
& \oplus_{i=1}^{\ell+\ell'-1} A_{2k_i+1} \oplus_{i=1}^{\ell'-1} A_{q_i}
\oplus C_{k_{\ell+\ell'}+1} \\
&& \\
(\ref{C8}) & \begin{array}{l}
C_{n+\ell+\ell'} \oplus p_1C_{\ell+\ell'}
\oplus_{i=1}^{\ell+\ell'-2} r_i C_{\ell+\ell'-i} \oplus \\
\oplus (k_1+1)A_{2\ell'-2} \oplus_{i=1}^{\ell'-2} \s_iA_{2\ell'-2i-2}
\end{array}
& \oplus_{i=1}^{\ell+\ell'} A_{p_i} \oplus_{i=1}^{\ell'-1} A_{2k_i+1}
\oplus C_{k_{\ell'}+1} \\
&& \\ \hline
\end{array}
\ee
}

\subsubsection{ \cw($D_N$,\ck) algebras}

\indent

As far as $D_N$ algebra are concerned, one can proceed in the same way as for
$B_N$ (replacing $B_M$ by $D_M$) except for the irregular embeddings
$B_k\oplus B_{N-k}\subset D_N$, that cannot be replaced by $D_M$ embeddings.
It is natural to think of them as related
to the particular cases $u=0$ or $u=1$ of the integral CDDs.
As these irregular
embeddings should not survive to the folding $D_N\ \rightarrow\ B_{N-1}$,
we focus on
$u=0$ and $\ell>1$ ($\ell=1$ is the principal CDD). Then, looking
at the $Sl(2)$ decomposition of the CDD (\ref{D4}),
we obtain $\underline{2N}=\cd_{n+\ell-1}\oplus
\cd_{\ell-1}$ which indeed corresponds to the reduction with respect to the
principal $Sl(2)$ of \footnote{We have used
the equality $2N=2n+2\ell-1+2\ell-1$, resulting from the decomposition
of the fundamental.} $B_{\ell-1}\oplus B_{N-\ell+1}$.
Note that the case $\ell=1$ could also be included in
the exceptions (with the convention $B_0=\emptyset $),
since the $Sl(2)$-principal embedding of $B_{N-1}$
in $D_N$ is also
the principal $Sl(2)$ of $D_N$. Thus the rules are
\bea
&&\mbox{CDDs } (\ref{D3}) \mbox{ and } (\ref{D4})
\ \longrightarrow\ B_{n+\ell-1}\oplus B_{\ell-1} \\
&& \mbox{other CDDs }\left\{\begin{array}{l}
\cd_p\ \longrightarrow\ D_p \\
\cd_{p/2}\ \longrightarrow\ A_p
\end{array}\right.
\ena
For the three $D_N$-CDDs ($u=0$, $u=1$, $u\geq2$) associated to one
$B_N$-CDD, the decomposition of the fundamental representation has
the same form, so that we will always (up to the above exceptions)
get the same type
of subalgebra \ck. However, because the tails have different grades
the $\cg_0$ subalgebras are different.
In the case where $u=1$, we get two $A_1$ subalgebras
from the tail, instead of the $D_{u+1}$ subalgebra arising when $u\geq2$.
But for $u=1$, the algebra $D_{u+1}=D_2$ is isomorphic
to $2A_1$, so that we include $u=1$ in the general treatment $u\geq2$.
Note that once more, it is the case $u=0$ that appears much different from
the other ones.

When $u=0$, we call $i_0$ the only
index such that
$p_{i_0}=0$ and $p_{i_0+1}=1$ (or $q_{i_0}=0$ and $q_{i_0+1}=1$):
it is the only index such that $r_{i_0}=1\neq0$ (or
$s_{i_0}=1\neq0$). In the case of integral grading we have $i_0=\ell-1$.

{\small
\be
\begin{array}{|c|c|c|}
\hline && \\
\mbox{CDD} & \ck \mbox{ in }\cw(D_N,\ck)
& \mbox{KM subalg. in \cw\ } (\mbox{up to }U(1)\mbox{ factors}) \\
&& \\ \hline && \\
\left.\begin{array}{l} (\ref{D1}) \\ (\ref{D2}) \end{array}\right\}
& D_{n+\ell-1}\oplus p_1D_{\ell-1}
\oplus_{i=1}^{\ell-2} r_i D_{\ell-i-1}
& D_{u+1}\oplus_{i=1}^{\ell-1} A_{p_i}\\
&& \\
\left.\begin{array}{l} (\ref{D3}) \\ (\ref{D4}) \end{array}\right\}
& B_{N-\ell+1} \oplus B_{\ell-1}
& (\ell-1) A_{p_{\ell-1}} \ \mbox{ with }\
p_{\ell-1}=\left\{\begin{array}{l} 1 \\ 0
\end{array}\right.\\
&& \\
(\ref{D5}) & (k_1+1) A_{2\ell+1} \oplus_{i=1}^{\ell} l_iA_{2\ell-2i+1}
& \oplus_{i=1}^{\ell+1} A_{2k_\ell+1} \\
&& \\
(\ref{D6}) & (k_1+1) A_{1}
& \ell A_{1} \\
&& \\
 \hline
\end{array}
\ee
}

{\small
\be
\begin{array}{|c|c|c|}
\hline && \\
\mbox{CDD} & \ck \mbox{ in }\cw(D_N,\ck)
& \mbox{KM subalg. in \cw\ } (\mbox{up to }U(1)\mbox{ factors}) \\
&& \\ \hline && \\
\left.\begin{array}{l} (\ref{D7}^+) \\ (\ref{D8}^+) \end{array}\right\}
& \begin{array}{l} (k_1+1)A_{2\ell-1}\oplus (q_1+1)D_{\ell} \oplus \\
\oplus_{i=1}^{\ell-1} \r_i A_{2\ell-2i-1}\oplus_{i=1}^{\ell-2}
s_i D_{\ell-i-1}
\end{array}
& \begin{array}{l} D_{u+1}\oplus A_{2k_\ell+1} \oplus \\
\oplus_{i=1}^{\ell-1} [A_{2k_i+1}\oplus A_{q_i}] \end{array}\\
&& \\
\left.\begin{array}{l} (\ref{D7}^-) \\ (\ref{D8}^-) \end{array}\right\}
& \begin{array}{l} (k_1+1)A_{2\ell-1}\oplus (q_1+1)D_{\ell-1} \oplus \\
\oplus_{i=1}^{\ell-1} \r_i A_{2\ell-2i-1}
\oplus_{i=1}^{\ell-2}  s_i D_{\ell-i-1}
\end{array}
& \begin{array}{l} D_{u+1}\oplus A_{2k_\ell+1} \oplus \\
\oplus_{i=1}^{\ell-1} [A_{2k_i+1}\oplus A_{q_i}] \end{array}\\
&& \\
(\ref{D9}^-)
& \begin{array}{l} (k_1+1)A_{2\ell-1}\oplus D_{\ell-1} \oplus \\
\oplus D_{\ell-i_0-1} \oplus_{i=1}^{\ell-1} \r_i A_{2\ell-2i-1}
\end{array}
& A_{2k_\ell+1}
\oplus_{i=1}^{\ell-1} [A_{2k_i+1}\oplus A_{q_i}] \\
&& \\
(\ref{D9}^+)
& \begin{array}{l} (k_1+1)A_{2\ell-1}\oplus D_{\ell} \oplus \\
\oplus D_{\ell-i_0-1} \oplus_{i=1}^{\ell-1} \r_i A_{2\ell-2i-1}
\end{array}
& A_{2k_\ell+1}
\oplus_{i=1}^{\ell-1} [A_{2k_i+1}\oplus A_{q_i}] \\
&& \\
\left.\begin{array}{l} (\ref{D10}) \\ (\ref{D11}) \end{array}\right\}
& \begin{array}{l} (p_1+1)D_{\ell} \oplus (k_1+1)A_{2\ell-3} \oplus \\
\oplus_{i=1}^{\ell-2} [r_i D_{\ell-i-1} \oplus \s_i A_{2\ell-2i-3}]
\end{array}
& D_{u+1}\oplus_{i=1}^{\ell-1} [A_{p_i}\oplus A_{2k_i+1}] \\
&& \\
(\ref{D12})
& \begin{array}{l} D_{\ell} \oplus (k_1+1)A_{2\ell-3} \oplus \\
 \oplus D_{\ell-i_0-1} \oplus_{i=1}^{\ell-2} \s_i A_{2\ell-2i-3}
\end{array}
& \oplus_{i=1}^{\ell-1} [A_{p_i}\oplus A_{2k_i+1}] \\
&& \\
\left.\begin{array}{l} (\ref{D13}) \\ (\ref{D14}) \end{array}\right\}
& \begin{array}{l}
(k_1+1) A_{2\ell-1}\oplus D_{n+\ell} \oplus q_1D_{\ell}\oplus\\
\oplus_{i=1}^{\ell-1} \r_i A_{2\ell-2i-1} \oplus_{i=1}^{\ell-2}
s_i D_{\ell-i-1}
\end{array}
& \begin{array}{l} D_{u+1} \oplus A_{2k_\ell+1} \oplus\\
\oplus_{i=1}^{\ell-1}[A_{2k_i+1}\oplus A_{q_i}]
\end{array}\\
&& \\
(\ref{D15}) & \begin{array}{l}
(k_1+1) A_{2\ell-1}\oplus D_{n+\ell} \oplus \\
\oplus D_{\ell-i_0-1} \oplus_{i=1}^{\ell-1} \r_i A_{2\ell-2i-1}
\end{array}
& A_{2k_\ell+1}
\oplus_{i=1}^{\ell-1}[A_{2k_i+1}\oplus A_{q_i}] \\
&& \\
\left.\begin{array}{l} (\ref{D16}) \\ (\ref{D17}) \end{array}\right\}
& \begin{array}{l}
D_{n+\ell+\ell'-1} \oplus_{i=1}^{\ell+\ell'-2}
r_i D_{\ell+\ell'-i-1} \oplus p_1D_{\ell+\ell'-1}\\
\oplus (k_1+1) A_{2\ell'-3} \oplus_{i=1}^{\ell'-2} \s_i A_{2\ell'-2i-3}
\end{array}
& D_{u+1} \oplus_{i=1}^{\ell+\ell'-1} A_{p_i}
\oplus_{i=1}^{\ell'} A_{2k_i+1} \\
&& \\
(\ref{D18})
& \begin{array}{l}
D_{n+\ell+\ell'-1} \oplus D_{\ell+\ell'-i_0-1} \oplus \\
\oplus (k_1+1) A_{2\ell'-3} \oplus_{i=1}^{\ell'-2} \s_i A_{2\ell'-2i-3}
\end{array}
&
\oplus_{i=1}^{\ell+\ell'-1} A_{p_i}
\oplus_{i=1}^{\ell'} A_{2k_i+1} \\
&& \\
\left.\begin{array}{l} (\ref{D19}) \\ (\ref{D20}) \end{array}\right\}
& \begin{array}{l}
(k_1+1) A_{2\ell+2\ell'-1} \oplus (q_1+1)D_{\ell'-1} \oplus \\
\oplus_{i=1}^{\ell'-2} D_{\ell'-i-1}
\oplus_{i=1}^{\ell+\ell'-1} \s_i A_{2\ell+2\ell'-2i-1}
\end{array}
& D_{u+1} \oplus_{i=1}^{\ell'-1} A_{q_i}
\oplus_{i=1}^{\ell+\ell'} A_{2k_i+1} \\
&& \\
(\ref{D21}) & \begin{array}{l}
(k_1+1) A_{2\ell+2\ell'-1} \oplus D_{\ell'-1} \oplus
 D_{\ell'-i_0-1} \\
\oplus_{i=1}^{\ell+\ell'-1} \s_i A_{2\ell+2\ell'-2i-1}
\end{array}
&
\oplus_{i=1}^{\ell'-1} A_{q_i}
\oplus_{i=1}^{\ell+\ell'} A_{2k_i+1} \\
&& \\
&& \\ \hline
\end{array}
\ee
}

\subsection{Duality between $B_N$ and $C_N$}

\indent

It has been proven in \cite{dual} that the \cw($B_N,B_N$) and $\cw(C_N,C_N)$
exhibit a duality, in the sense that the Abelian Toda model constructed on
$C_N$ is obtained from the Abelian Toda model built on $B_N$ by replacing the
roots system by its dual, and the coupling constant $\beta$ by
$\frac{-1}{\beta}$.

Using the CDDs, it is possible to see which of the (other)
\cw($B_N$,\ck) and \cw($C_N,\widetilde{\ck}$) algebras can be dual.
Indeed, using the
duality between $B_N$ and $C_N$ DDs that consists in exchanging the colors
black
and white in the nodes, one obtains the following candidates for the
"B-C duality"
\bea
&& \uo{2k_1+1}\tr\bu\tr\uo{2k_2+1}\tr\bu\tr\uo{2k_3+1}\tr\pol\tr\bu
\tr\uo{2k_{\ell-1}+1}\tr\bu\tr\uo{k_\ell}\td\bz \label{BC1}\\
&& \bu\tr\bu\tr\pol\tr\bu\tr\bu=\bu \label{BC2}\\
&& \uo{2k_1+1}\tr\bd\tr\uo{2j_1+1}\tr\bd\tr\uo{2k_2+1}
\tr\bd\tr\uo{2j_2+1}\tr\pol
\tr\bd\tr\uo{2k_\ell+1}\tr\bd\tr\uo{j_\ell}\td\bz \label{BC3}\\
&& \und{\bz\tr\po\tr\bz}_{2k_1+1}\tr\bd\tr\und{\bz\tr\po\tr\bz}_{2j_1+1}
\tr\bd\tr\und{\bz\tr\po\tr\bz}_{2k_2+1}\tr\bd\tr
\und{\bz\tr\po\tr\bz}_{2j_2+1}\tr\pol\tr\bd\tr
\und{\bz\tr\po\tr\bz}_{2j_{\ell-1}+1}\tr\bd\tr
\und{\bz\tr\po\tr\bz}_{k_\ell}\td\bz\  \label{BC4}\\
&& \puo{2k_1+1}\tr\bu\tr\pol\tr\bu\tr\puo{2k_{\ell+1}+1}
\tr\bd\tr\puo{2j_1+1}
\tr\bd\tr\puo{2k_2+1}
\tr\bd\tr\pol\tr\bd\tr\puo{2j_{\ell'-1}+1}
\tr\bd\tr\puo{k_{\ell+\ell'}}\td\bz
\label{BC5}\\
&& \und{\bu\tr\po\tr\bu}_n\tr\bd\tr\uo{2k_1+1}\tr\bd\tr\uo{2j_1+1}
\tr\bd\tr\uo{2k_2+1}\tr\bd\tr\uo{2j_2+1}\tr\pol\tr\bd\tr\uo{2k_\ell+1}
\tr\bd\tr\uo{j_{\ell}}\td\bz
\label{BC6}\\
&& \und{\bz\tp\po\tp\bz}_{2k_1+1}\tr\bu\tr\und{\bz\tp\po\tp\bz}_{2k_2+1}
\tr\bu\tr\pol\tr\bu\tr\und{\bz\tp\po\tp\bz}_{2k_{\ell+1}+1}\tr\bd\tr
\und{\bz\tp\po\tp\bz}_{2j_1+1}\tr\bd\tr\und{\bz\tp\po\tp\bz}_{2k_{\ell+2}+1}
\tr\bd\tr\und{\bz\tp\po\tp\bz}_{2j_2+1}\tr\pol\tr
\und{\bz\tp\po\tp\bz}_{2k_{\ell+\ell'}+1}\tr\bd\tr
\und{\bz\tp\po\tp\bz}_{j_{\ell'}}\td\bz\nonumber\\
\label{BC7}
\ena
Notice that one has to pay attention, when comparing $B_N$ and $C_N$
CDDs, to the difference between $\ell'=\ell$ and $\ell'=\ell-1$. For
instance, (\ref{B6}) seems very similar to (\ref{C6}), but has one piece
of length $q$ more than the other: to get (\ref{BC6}), one has to take
(\ref{B6}) with $q_i$ odd $(i\neq\ell)$ on the one hand, but
(\ref{C8}) with $\ell=0$, $p_i$ odd $(i\geq2)$ and $p_1=0$ on the other
hand.

\indent

Looking at the fundamental representation, it is easy to
see that in most of the cases we have the same decomposition for
the $B_N$ and $C_N$ algebras (except one $\cd_0$ that allows to go
from $\underline{2N}$ to $\underline{2N+1}$).
This means that the spin contents
of the corresponding $W$ algebras will be different, since it is
obtained by a symmetric (antisymmetric) product of the fundamental
by its contragredient for $C_N$ ($B_N$). The exceptions
that have different fundamental
decomposition for $B_N$ and $C_N$ are (\ref{BC2}) and (\ref{BC6}).
The principal CDD (\ref{BC2}) lead to the well-known algebras
\cw($B_N,B_N$) and \cw($C_N,C_N$), and a direct calculation of
the adjoint decomposition for (\ref{BC6}) shows that the spin
content is still different.

Thus, the only $W$ algebras related by a "B-C duality" are
\cw($B_N,B_N$) and \cw($C_N,C_N$). It is nevertheless possible
that the $W$ algebras pointed out in this paragraph are related
through another relation weaker than this duality.

\sect{Conclusion}

\indent

In this paper, we have presented a classification of CDDs for A,B,C,D
algebras, and connected them to the classification of \cw(\cg,\ck)
algebras. The CDD may be viewed as a kind of Dynkin diagram for these
algebras, so that one can wonder whether it is still true for a more
general class of $W$ algebras. For such a purpose, one can think of
enlarging the definition of CDDs by allowing the grades to take any
values and try to connect $W$ algebras with "generalized" CDDs in a
different way than the $Sl(2)$ classification used here. If it is
possible, this will imply that a "generalised"
CDD encode all the relevant informations
concerning the $W$ algebras (as it is the case for CDDs with \cw(\cg,\ck)
algebras). Note however that a generalised CDD will always
be related to a Cartan generator of \cg\ (but not of a $Sl(2)$
\cg-subalgebra), through the grades of the simple roots.

Finally, let us remark that a similar treatment for super-$W$ algebras,
with some "super-CDDs" classifying $OSp(1|2)$ sub-superalgebras, may
perhaps be done. One has first to know if the $OSp(1|2)$ embeddings can
be classify by such diagrams: to our knowledge, this has not yet
been studied, but the notion of super-defining vectors has already being
introduced in \cite{nous}.

\indent

{\bf \large{Acknowledgements}}

\indent

I would like to thank Jens Ole Madsen, Jens Lyng Petersen and Paul Sorba
for reading this manuscript.

%\newpage

\end{document}